\documentclass[aip,amsmath,amssymb,reprint,]{revtex4-1}

\usepackage{graphicx}
\usepackage{dcolumn}
\usepackage{bm}
\usepackage[utf8]{inputenc}
\usepackage[T1]{fontenc}
\usepackage{mathptmx}
\usepackage[dvipsnames]{xcolor}

\begin{document}

\preprint{AIP/123-QED}

\title[Practicality of InSb as a magneto-optical platform]{Indium Antimonide -- constraints on practicality as a magneto-optical platform for topological surface plasmon polaritons}

\author{S. Pakniyat}
\email{pakniyat@uwm.edu}
\affiliation{Department of Electrical Engineering, University of Wisconsin-Milwaukee, Milwaukee, Wisconsin 53211, USA}%

\author{Y. Liang}%
 \email{yi.liang.nankai@gmail.com}
\affiliation{ Department of Physics and Astronomy, West Virginia Universityh, Morgantown, West Virginia 26506, USA}%

\affiliation{ Guangxi Key Lab for Relativistic Astrophysics, Center on Nanoenergy Research, School of Physical Science and Technology, Guangxi University, Nanning, Guangxi 530004, China}%

\author{Y. Xiang}
\email{yinxiaoermao@googlemail.com}
\affiliation{ Department of Physics and Astronomy, West Virginia Universityh, Morgantown, West Virginia 26506, USA}%

\author{C. Cen}
\email{cheng.cen@mail.wvu.edu}
\affiliation{ Department of Physics and Astronomy, West Virginia Universityh, Morgantown, West Virginia 26506, USA}%

\author{J. Chen}
\email{JUC48@pitt.edu}
\affiliation{Department of Electrical and Computer Engineering, and Petersen Institute of Nano Science and Engineering, University of Pittsburgh, Pennsylvania 15261, USA}

\author{G. W. Hanson}
\email{george@uwm.edu}
\affiliation{Department of Electrical Engineering, University of Wisconsin-Milwaukee, Milwaukee, Wisconsin 53211, USA}

\date{\today}
            
\begin{abstract}
Magnetic-field-biased indium antimonide (InSb) is one of the most widely-discussed materials for supporting nonreciprocal surface plasmon polaritons (SPPs), which have recently been shown to be topological. In this work, we provide a critical assessment of InSb as a magneto-optical SPP platform, and show that it is only viable under a narrow set of conditions. 
\end{abstract}

\maketitle

\section{\label{sec:level1}Introduction}
Continuous media with broken time-reversal symmetry, such as a magnetized semiconductor, have recently been shown to be topologically nontrivial \cite{davoyan}, \cite{Mario2} \cite{Mario3} \cite{MH1}. Topological surface waves have unique optical properties, namely, one-way propagation (immunity to back-scattering) and, since they exist in the bulk bandgap, upon encountering discontinuities they do not diffract into the bulk \cite{Soljacic2014} \cite{rechtsman2013photonic}. Topological effects in photonic systems are promising in the realization of devices \cite{Haldane2} such as optical isolators and circulators. 

Theoretical/simulation studies of nonreciprocal SPPs, including topological SPPs, often cite InSb as an example of a material providing the necessary gyrotropic permittivity tensor \cite{airSPP} \cite{PRA} \cite{Villa15} \cite{Fan16} \cite{Boyd} \cite{Ali18leaky} \cite{Villa2} \cite{Hu19} \cite{Zhang19} \cite{aliPWG}. However, here we show that, while SPPs with reasonable propagation characteristics can be obtained, there exists severe constraints that limit performance. In particular, we find that for InSb to serve as a viable SPP platform under modest bias field strengths, one needs to use undoped materials and low, but not too-low, temperatures to obtain sufficiently-high mobility and a reasonably-sized bulk bandgap. 
 
In the following, there is a brief review on topological SPP properties in a dissipation-less system in the Voigt configuration, in which the propagation vector is
perpendicular to the in-plane magnetic bias, and there are topological SPPs crossing the bulk bandgaps of the gyrotropic plasma medium. Next, using time-domain spectroscopy (THz-TDS), the measured reflection from a magnetized InSb crystal is considered. Using the measured parameters at different temperatures, the properties of topological SPPs in a realistic plasma material are examined. It is shown that temperature plays an important role in optimizing the SPP propagation properties. Lastly, with the help of a symmetric grating launcher, the SPPs are observed at the interface of gold/InSb at various temperatures using a far-field measurement. The measured SPP resonance frequencies are consistent with the values theoretically estimated. 

\section{\label{sec:level2}Electromagnetic Model and Results}
\subsection{\label{sec:level3}Material Model and Bulk Modes}
\bigskip Consider a half-space plasma medium (InSb) with unit normal vector $\hat{z}$, biased by an in-plane external
magnetic field $\mathbf{B}_{0}=\widehat{\mathbf{y}}B_{0}.$ The gyrotropic InSb can be characterized by a simplified Drude model with a
dielectric tensor in the form \cite{Palik} $\mathbf{\bar{\varepsilon}}_{r}%
=\varepsilon_{t}\mathbf{\bar{I}}_{t}+i\varepsilon_{g}(\widehat{\mathbf{y}%
}\mathbf{\times\bar{I})+}\varepsilon_{a}\widehat{\mathbf{y}}\widehat
{\mathbf{y}}$ (assuming the time harmonic variation $e^{-i\omega t}$) where
\begin{align}
\varepsilon_{t}  & =\varepsilon_{\infty}-\frac{\omega_{p}^{2}(1+i\Gamma
/\omega)}{(\omega+i\Gamma)^{2}-\omega_{c}^{2}},\text{ \ }\varepsilon
_{a}=\varepsilon_{\infty}-\frac{\omega_{p}^{2}}{\omega(\omega+i\Gamma)}, \text{ \ }\nonumber\\
\text{ \ }\varepsilon_{g}  & =\frac{\omega_{c}\omega_{p}^{2}}{\omega\left[
\omega_{c}^{2}-(\omega+i\Gamma)^{2}\right]  },\label{e2}%
\end{align}
and where $\varepsilon_{\infty}$ is the background permittivity (high-frequency
dielectric constant). The plasma frequency is $\omega_{p}=$ $\sqrt{n_{e}q_{e}^{2}/(m^{\ast}%
\varepsilon_{0})}$, $\omega_{c}=$ $-q_{e}B_{0}/m^{\ast}$ is the cyclotron frequency, and $\Gamma
=-q_{e}/\mu m^{\ast}$ is the collision frequency. Also, $n_{e}$ is the free electron density, $q_{e}=-e$ is the electron charge, $\varepsilon_{0}$ is the free-space permittivity, $m^{\ast
}=m_{\text{eff}}$ $m_{0}$ is the effective electron mass, $m_{0}$ is the vacuum electron mass, and $\mu$ is the mobility. In the absence of a
magnetic bias, the gyrotropic plasma turns to a dispersive isotropic
medium having permittivity $\varepsilon_{a}.$  Here, we ignore the phonon
contribution in the material model, which is an accurate assumption at
frequencies below the optical phonon resonances (and at low temperatures). There are also contributions from heavy and light hole bands \cite{P2} but both were found to be negligible; the former because of the large mass and the latter due to small carrier density, even at room temperature. The plasma frequency $\omega_p$ sets the frequency scale for the occurrence of bandgaps, and we need a sufficient magnetic bias so that the cyclotron frequency $\omega_c$ is not too small compared to $\omega_p$ in order to achieve a usual amount of nonreciprocity.
\begin{figure}
  \includegraphics[width=1.0\columnwidth]{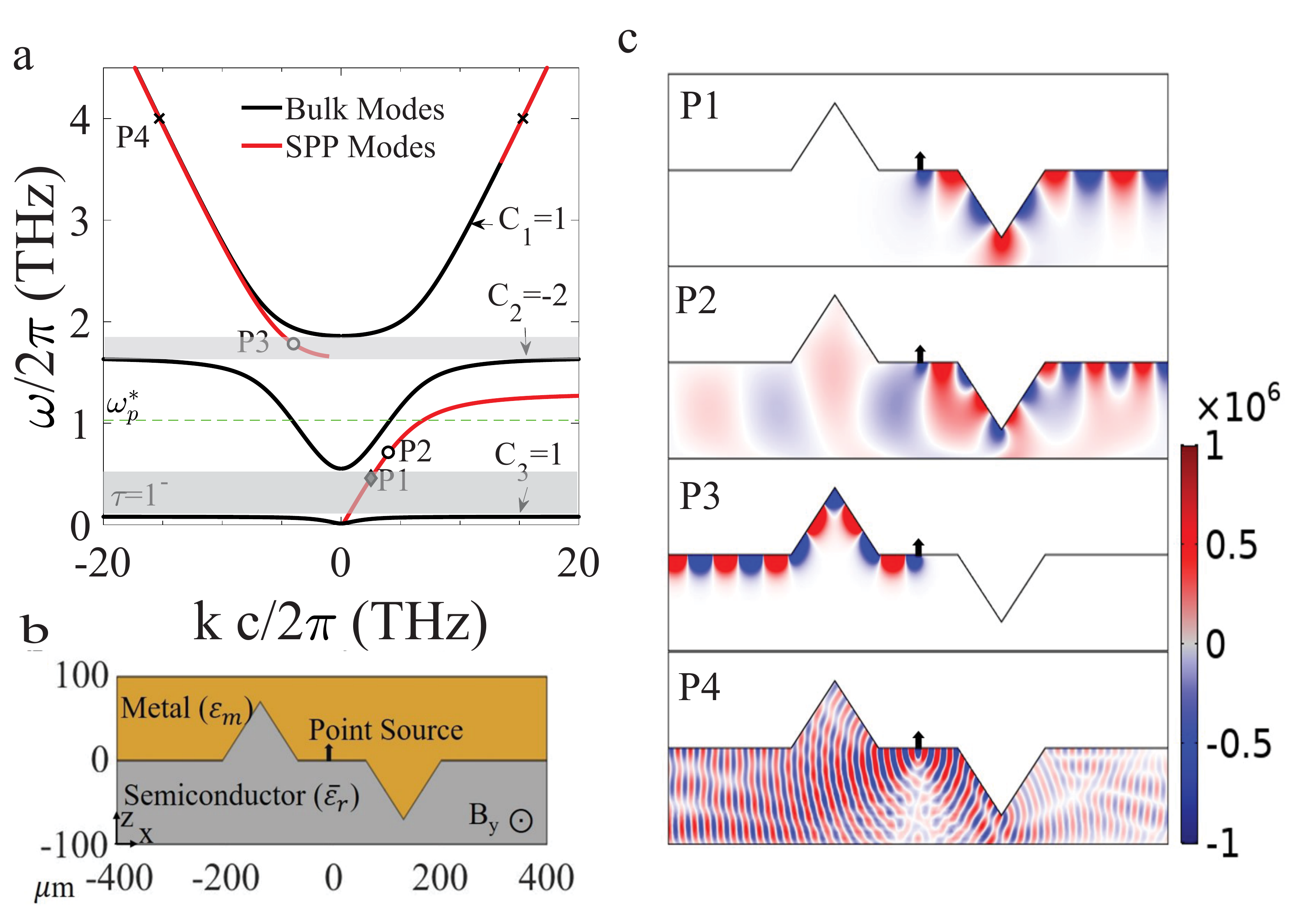}
  \caption{Bulk modes and surface plasmons for a
metal-InSb (idealized) interface. (a) Bulk and SPP dispersion for dissipation-less InSb characterized by (\ref{e2}) with parameters $n_e=3\times
10^{21}\text{m}^{-3},$ $m^{\ast}=0.015m_{0}$, $\varepsilon_{\infty}=15.68$, and mobility $\mu=\infty$ ($\omega_p^{\ast}=\omega_p/\sqrt{\epsilon_{\infty}}=2\pi(1.014 \text{THz})$). The carrier densities and other numerical values have been obtained from fitting (1) to measurements \cite{InSb}. The in-plane magnetic field is $B=0.7$T ($\omega_c=1.28\omega_p^{\ast}=2\pi(1.3 \text{THz})$), and the metal permittivity is $\varepsilon
_{m}=-10^{4}$. The gray regions are the magnetic-field dependent bandgaps (b) The geometrical schematic of the structure under study.
SPPs are excited by a dipole source located at the metal-InSb interface. (c) The
electric field distribution ($E_{z}$) of SPPs at different frequencies P1-P4, $0.5, 0.7, 1.76$ and $4$ THz, respectively, marked in (a). Points P1 and P3 are points on the SPP dispersion where the SPPs are topologically protected.
  }
 \label{DISP}
\end{figure}
The characteristics of the bulk modes in a gyrotropic medium depend on
the direction of propagation with respect to the external magnetic bias. In
the well-known Voigt configuration, the bulk modes are propagating
perpendicular to the magnetic bias. In this case, the bulk waves can be
decoupled into a TE (with electric field along the magnetic bias) and a TM
(with electric field in a plane perpendicular to the bias) wave
described by $k_{\text{TE}}^{2}=\varepsilon_{\text{a}}k_{0}{}^{2}$ and
$k_{\text{TM}}^{2}=\varepsilon_{\text{eff}}k_{0}^{2}$, respectively, where
$\varepsilon_{\text{eff}}=(\varepsilon_{t}^{2}-\varepsilon_{g}^{2}%
)/\varepsilon_{t}$ and $\ k_{0}$ is the free space wave number. 
Figure \ref{DISP}a (black lines) shows
the dispersion diagram of TM bulk modes propagating in dissipation-less InSb. There are two magnetic-field dependent bandgaps. Each TM band is characterized by a non-zero integer Chern number $C_{n}$, an intrinsic property of the bulk band structure, so that Chern numbers sum to zero, as expected \cite{Han1}. The gap Chern number, that is, the sum of the Chern numbers below the gap, is non-zero,
indicating the number of topologically protected surface modes crossing the band gaps. The TE mode, with $C_{n}=0$, is topologically-trivial and won't be discussed further. Here, in the material model non-locality is ignored except to provide a momentum cut-off \cite{Mario2}. The effect of non-locality
is evident only for very large wavenumbers \cite{Shi}, in which case the backward waves vanish for realistic levels of loss \cite{Ali}. 

\subsection{Surface Plasmon Polaritons}
Despite the non-reciprocal nature of the medium itself, in the Voigt configuration the bulk
dispersion behavior is reciprocal, although an interface
will break this reciprocity. The TM-SPP dispersion equation is \cite{davoyan,kush}%
\begin{equation}
\frac{\gamma_{zm}}{\varepsilon_{m}}+\frac{\gamma_{z}}{\varepsilon_{\text{eff}%
}}=\frac{\varepsilon_{g}k_{x}}{\varepsilon_{t}\varepsilon_{\text{eff}}},%
\end{equation}
where $\gamma_{z}=\sqrt{k_{x}^{2}-k_{0}^{2}\varepsilon_{\text{eff}}}$, $\gamma_{zm}=\sqrt{k_{x}^{2}-k_{0}^{2}\varepsilon_{m}},$ and
$\varepsilon_{m}$ is the effective permittivity of the top (metal) layer. The SPP is
propagating on the $x-y$ plane (interface of two media) with the propagation
constant $k_{x}$. If isotropic materials form the interface, for an SPP to propagate the two permittivities must have opposite signs. In the gyrotropic plasma-isotropic metal case, it can be shown analytically that in the bandgaps, where $\varepsilon_{\text{eff}}<0$, for an SPP to exist we need $\varepsilon_{t}>0$. Outside of the bandgaps it can be seen numerically that no SPP exists when $\varepsilon_{t}<0$.

Figure \ref{DISP}a shows the dispersion of the SPP at the interface of a dissipationless gyrotropic semiconductor and an opaque medium (red lines), with the interface geometry depicted in Fig. \ref{DISP}b. Figure \ref{DISP}c shows the electric field profile of SPP propagation at different points on the dispersion curve, P1-P4, shown in Fig. \ref{DISP}a, obtained from a full wave simulation using COMSOL assuming a dipole source. As discussed above, SPPs crossing the bandgaps, for example at frequency points P1 and P3, are topologically protected surface
waves, meaning they are unidirectional and propagate along the surface
without reflection. In the frequency range between the two bandgaps, although the dispersion is strongly nonreciprocal, upon reflection, surface wave can couple into the bulk modes, see, e.g., point P2. At higher frequencies, for example at point P4, SPPs are bi-directional; the right- and left-going SPPs have approximately the same momentum, $\omega(k)=\omega(-k)$. Partial reflection of the wave occurs, since at this frequency the material itself allows propagation in both directions.

\subsection{Realistic InSb model}
The previous discussion was for lossless materials. In the following, the properties of bulk and topological SPPs for a realistic InSb model are examined. We consider that for SPP applications we require, or it is at least very desirable, that 1) ${L_\textrm{SPP}} /{\lambda_\textrm{SPP}}\gg 1$, where SPP propagation length is $L_{\text{SPP}}=1/(2\operatorname{Im}\left(  k_{\text{SPP}}\right))$ and SPP wavelength is $\lambda_{\text{SPP}}=2\pi/\operatorname{Re}\left(  k_{\text{SPP}}\right)$, 2) SPPs that are nonreciprocal, so that they are backscattering-immune, and 3) SPPs exist in the bulk bandgaps, so that they will not diffract into the bulk upon encountering a discontinuity. To meet these criteria, the requirements on the material properties are very stringent.

In the experiment, we use an InSb crystal from the manufacturer MTI Corporation \cite{MTI} having dimensions $10\times10\times0.5$ mm, undoped, with one side polished. In \cite{InSb}, we have extracted the carrier density, mobility, and effective mass of InSb at various temperatures from $5^\circ$K to $300^\circ$K and under external bias fields up to $0.7$T, determined by far-field time-domain terahertz spectroscopy (THz-TDS) in the range 0.5–3 THz. Four material samples were tested, and the results below represent values from one of the samples (two others were similar, and one showed poorer performance). For the metal, we assume $\varepsilon_{m}=-800+i550$ \cite{ND}. If the metal is less lossy, longer propagation lengths will be obtained. Also, note that $L_{\text{SPP}}$ was not directly measured, but was calculated based on the measured material parameters.
\begin{figure} 
	\includegraphics[width=1.0\columnwidth]{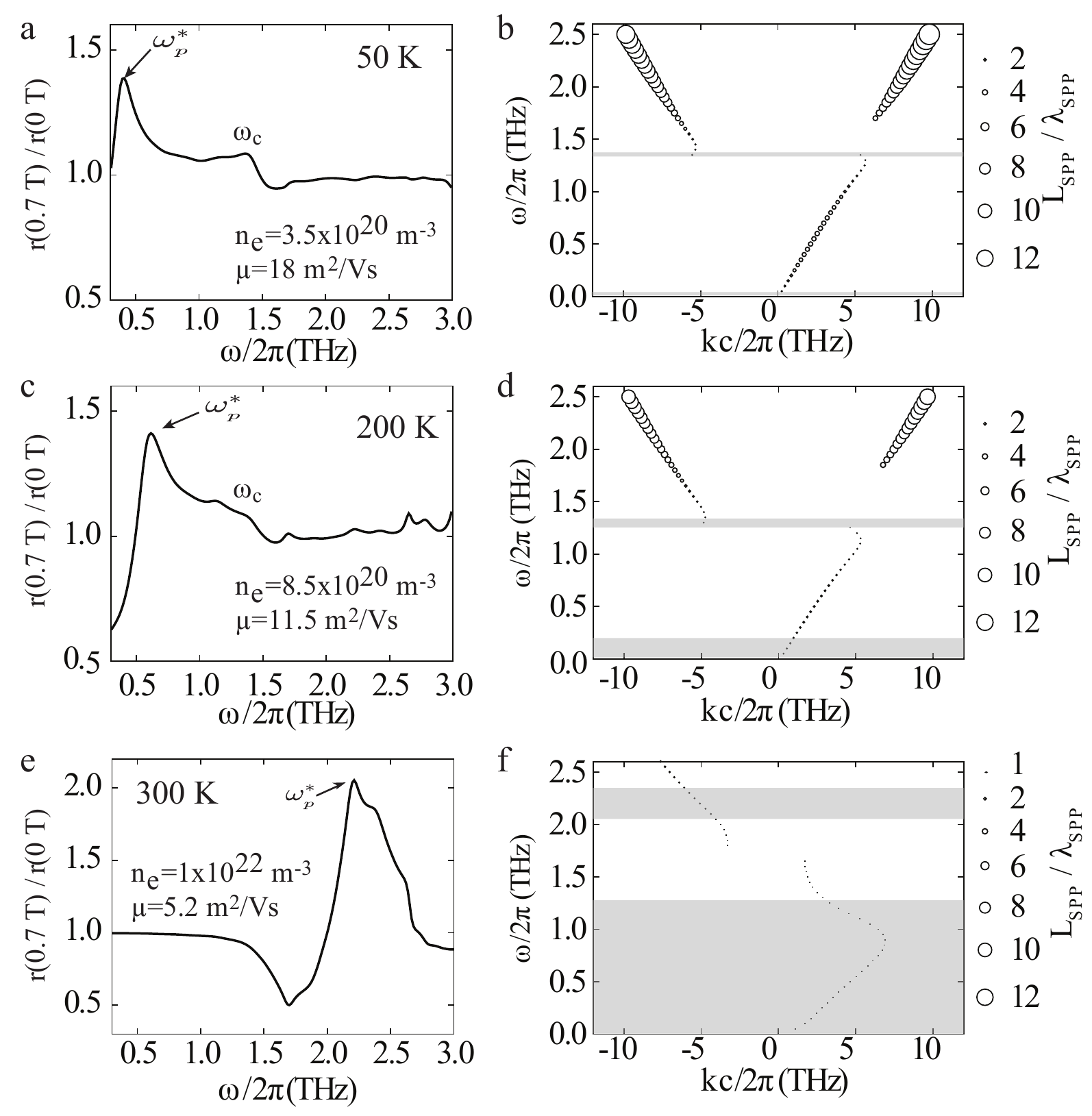}
  \caption{(a,c,e) Experimentally measured reflectance spectra of the InSb sample at different temperatures. The values for $n_e$ and $\mu$ in each panel are the extracted InSb parameters by fitting measurement data to the Drude model. (b,d,f) The SPP dispersion diagrams and SPP propagation length (indicated by the size of the circles). Shaded regions indicate the bandgaps. }
 \label{bare}
\end{figure}

Figure \ref{bare} a,c,e shows the reflectance spectra, $\tilde{R}=r(B)/r(0)$, of an air/InSb sample measured at different temperatures for $B=0.7$ T. The material parameters can be obtained using the Drude model and the analytical reflection coefficient in the Voigt configuration defined as
\begin{equation}
R_{TM}=\frac{\sqrt{\varepsilon _{\text{eff}}}-\sqrt{\varepsilon _{d}}}{\sqrt{%
\varepsilon _{\text{eff}}}+\sqrt{\varepsilon _{d}}}
\label{NRE}
\end{equation}%
where $\varepsilon _{d}$ is the permittivity of the isotropic material (here 
$\varepsilon _{d}=1)$ and $r=\left\vert R_{TM}\right\vert ^{2}$.
As shown in Fig. \ref{bare}a,c,e, the carrier concentration decreases as the temperature decreases, resulting in
shifting the reduced edge plasma frequency, $\omega_p^{\ast}=\omega_p/\sqrt{\epsilon_{\infty}}$, toward lower frequencies. In the bandgap the transparency of the biased crystal increases, causing the second peak in the reflectance spectrum, which is a function of cyclotron frequency, plasma frequency, and magnetic bias. As temperature
decreases down to $50^\circ$K, the mobility increases as expected since the scattering rate increases with temperature. 

Based on the measured material parameters, the SPP dispersion diagram and propagation length of the SPP excited at the interface of a gold/InSb interface at each temperature are shown in Fig. \ref{bare}b,d,f. At
lower temperatures the propagation length is larger due to the higher mobility, although the bandgaps become very narrow. 

\begin{figure}
  \includegraphics[width=1.0\columnwidth]{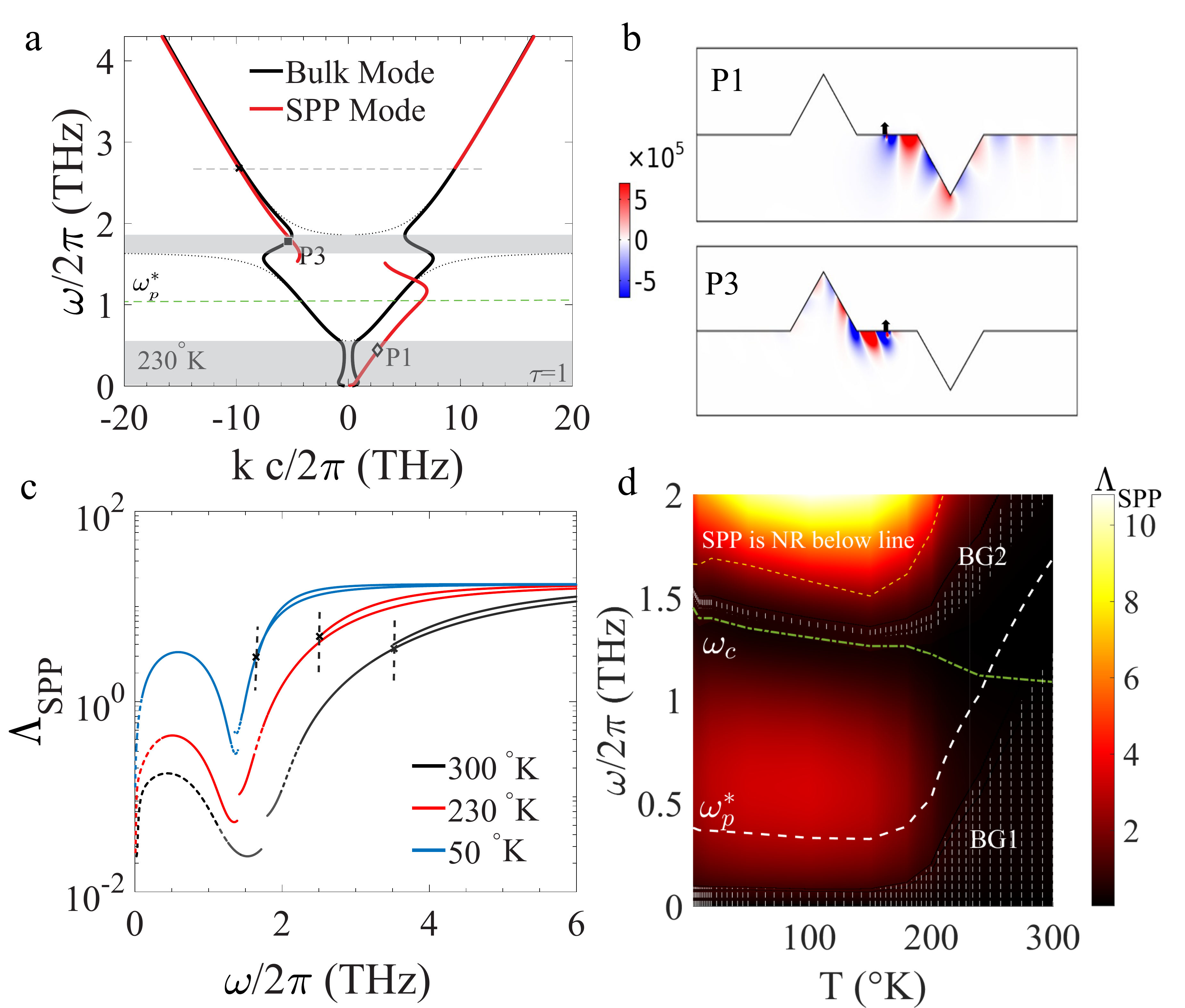}
  \caption{Bulk modes and nonreciprocal InSb-metal interface SPPs for a realistic InSb model with finite dissipation, (a-b) bulk and SPP dispersion diagrams using InSb material parameters at $T=230^\circ$K, which are $n_e=3\times
10^{21}\text{m}^{-3}$, $\mu=9 \text{m}^2/\text{Vs}$, $m^{\ast}=0.015m_{0}$, $\varepsilon_{\infty}=15.68$, extracted from measurement \cite{InSb}, and $B=0.7$T, $\varepsilon
_{m}=-800+i550$. Solid lines correspond to finite dissipation and dotted lines correspond to the infinite mobility cases. Below the dashed horizontal line the SPP is nonreciprocal (NR), and sometimes unidirectional (b) The electric field distribution of the unidirectional SPP, at two resonance frequencies P1 and P2, excited by a point source located at the interface for the finite-mobility case. (c) SPP propagation length ($L_\textrm{SPP}$) in a dissipative system at three different temperature, where $\Lambda_\textrm{SPP}$=${L_\textrm{SPP}} /{\lambda_\textrm{SPP}}$. To the left of the dashed vertical lines the SPP is nonreciprocal. The dashed sections indicate the SPP is within the bandgaps. (d) Density plot of normalized SPP propagation length, $\Lambda_\textrm{SPP}$, versus frequency and temperature. Shaded regions indicate the bandgaps.     
  }
 \label{InSb}
\end{figure}

Figure \ref{InSb}a shows the bulk and SPP dispersion for InSb at $230^\circ$K. Loss leads to modified bulk plasmon dispersions, in which case there no longer exists a true bandgap (comparing the solid black lines in Fig. \ref{InSb}a with the dotted black lines, which indicate the dissipation-less case from Fig. \ref{DISP}a). However, there are still distinguishable regions of dispersion.
 
The field profile of the SPPs at two dispersion points P1 and
P2 (Fig. \ref{InSb}a) are shown in Fig. \ref{InSb}b. In a dissipative system, nonreciprocal (unidirectional) SPPs are still immune to back-scattering upon encountering a discontinuity. Figure \ref{InSb}c
shows the propagation length of SPPs supported by InSb at three different temperatures. At high temperature the mobility is low ($\mu=5.2$ $\text{m}^{2}/\text{Vs}$), so that the propagation length of the one-way SPP (at frequencies below the dashed line, where SPPs are nonreciprocal) is a small fraction of the SPP wavelength (Fig. \ref{InSb}c, black curve). By reducing the temperature, the bandgap range decreases but the mobility increases, resulting in longer propagation lengths as shown in Fig. \ref{InSb}c, blue curve. In each case, only to the left of the dashed vertical lines the SPP is nonreciprocal. Moreover, only along the dashed sections is the SPP within the bandgaps. 

In Table \ref{Lspp} the values for the bandgap frequency range and the propagation length of the nonreciprocal SPPs at different frequencies, in and between the two bandgaps, and at various temperatures, are listed for $B=0.7$T. As shown, there is a trade-off between the propagation length and the size of the bandgap -- as temperature is lowered, the bandgap can become quite narrow and impractical. Therefore, temperature needs to be carefully chosen, low enough to yield sufficient SPP propagation lengths yet not too low so that the bandgap is wide-enough to work within. Furthermore, unless magnetic bias is adjusted appropriatly, the SPP will not exist within the bandgap.  

Given that one may consider operation within the bandgap as most desirable, for this bias, only for the lowest temperature is ${L_\textrm{SPP}} /{\lambda_\textrm{SPP}}\approx 1$, although longer propagation lengths are found between the two bandgaps. Similar to Table \ref{Lspp}, Fig. \ref{InSb}d shows a density plot of SPP propagation length $\Lambda_\textrm{SPP}$=${L_\textrm{SPP}} /{\lambda_\textrm{SPP}}$ versus frequency and temperature. The temperature--bandgap width tradeoff is prominent, as are the relatively short values of non-reciprocal SPP propagation length except between the bandgaps.  
\begin{table}
\caption{SPP propagation length and bandgap width in frequency at various temperatures. BG$_1$ (which starts at 0 THz; the provided value is for the top of the bandgap) and BG$_2$ are the lower and upper bandgaps, respectively. BTWN-BGs indicates at the midpoint between the two bandgaps (in the bulk passband), $\epsilon_m=-800+i550$ \cite{ND}, $B=0.7$T, and $\Lambda_\textrm{SPP}$=${L_\textrm{SPP}} /{\lambda_\textrm{SPP}}$.}
\label{Lspp}
\begin{ruledtabular}
\begin{tabular}[c]{lllllll}%
Temperature ($ ^\circ $K)&              300 & 250    & 200   & 150   & 100  & 50\\ \hline \hline
BG$_1$ (THz)&                           1.23 & 0.72  & 0.19  & 0.08  & 0.08 & 0.08 \\ \hline
BG$_2$ (THz)&                           2.01-& 1.60- & 1.33- & 1.30- & 1.35-& 1.39-\\ 
            &                           2.32 & 1.84  & 1.41  & 1.34  & 1.38 & 1.43 \\ \hline
$\Lambda$ in-BG$_1$&                    0.17 & 0.30  & 0.76  & 0.91  & 1.03 & 1.00\\
$\Lambda$ BTWN-BGs&                     0.03 & 0.25  & 1.20  & 2.90  & 3.30 & 3.30 \\
$\Lambda$ in-BG$_2$&                    0.46 & 0.48  & 0.50  & 0.75  & 0.66 & 0.61\\
\end{tabular}
\end{ruledtabular}
\end{table}
Table \ref{doped} provides a comparison of SPP properties between undoped, N-doped, and P-doped samples at $T=77^\circ$K. For the undoped case, material parameters values were taken from measurements \cite{InSb}, whereas for the doped cases we used parameter values from the manufacturer website \cite{MTI}. The P-doped case has very low mobility and poor SPP properties, and won't be discussed further. The undoped case discussed above provides good SPP propagation below 2 THz and $T \lessapprox 200^\circ$K. Because of the large carrier density in the N-doped case, the plasma frequency, and, hence, the upper bandgap, occurs at higher frequency than for the undoped case, around 10THz for $B=0.7$T. However, for this strength bias, $\omega_c/\omega_p$ is small, and poor SPP propagation is obtained. In order to obtain comparable SPP performance for the N-doped material, one would need $B=15$T, which is difficult to obtain, at which point the higher bandgap occurs near 30 THz (here we assume the mobility is the same as the low-THz values). 
\begin{table}
\caption{Comparison of plasmon propagation properties in the undoped, N-doped and P-doped crystals at $T=77^\circ$K, considering $m^{\ast}=0.015m_0, \epsilon_{\infty}=15.7$ and $\epsilon_m=-800+i550$. $\Lambda$=${L_\textrm{SPP}} /{\lambda_\textrm{SPP}}$. Dashes indicate no SPP exists. BG$_1$ starts at 0 THz; the provide value is for the top of the bandgap.}
\label{doped}

\begin{ruledtabular}
\begin{tabular}[c]{ll|ll|l}%
$T=77^{\circ}$K                    & undoped   &  \multicolumn{2}{c|}{N-doped} & P-doped \\ \hline 
$n_e\times10^{21}(\text{m}^{-3})$  & \ 0.32    &  \multicolumn{2}{c|}{350}     & \ 35    \\ \hline 
$\mu (\text{m}^2/\text{Vs}$)       & \ 17      &  \multicolumn{2}{c|}{4.5}     & \ 0.2   \\ \hline 
$\omega_p^{\ast}/2\pi (\text{THz})$& \ 0.33    & 10.9      & \ \ 10.9          & \ 3.4   \\ \hline
B (T)                              & \ 0.7     & 0.7       & \ \ 15            & \ 0.7   \\ \hline
$\omega_c/\omega_p^{\ast}$         & \ 3.9     & 0.12      & \ \ 2.5           & \ 0.37  \\ \hline
$\Gamma/\omega_p$                  & \ 0.08    & 0.01      & \ \ 0.01          & \ 0.7   \\ \hline
BG$_{1}$ (THz)                     & \ 0.08    & 10.3      & \ \ 3.70          & \ 2.9   \\ \hline
BG$_2$ (THz)                      & \ 1.35-   & 11.0-     & \ \ 30.0-         & \ 3.7   \\
                                   & \ 1.38    & 11.6      & \ \ 31.7          & \ 4.2   \\ \hline
$\Lambda$  \ in-BG1                & \ 1.0     & 0.02      & \ \ 1.6           & \ 0.2   \\ 
$\Lambda$  \ BTWN-BGs              & \ 3.3     & -         & \ \ 3.3           & \ -     \\ 
$\Lambda$  \ in-BG2                & \ 0.65    & 0.5       & \ \ 1.1           & \ -     \\ 

\end{tabular}
\end{ruledtabular}
\end{table}

In summary, one may conclude that for working at low THz ($\lessapprox$ 2THz) and moderate bias strength ($\lessapprox$1T) the undoped material is the only viable option. Room temperature operation is not feasible. For the N-doped material, aside from $B=15$T, below $\lessapprox$ 5T, there is no SPP between the two bandgaps, but for biases levels such as $B=5,7.5,12$T  the maximum ${L_\textrm{SPP}} /{\lambda_\textrm{SPP}}$ is 0.4, 1.1 and 2.3, respectively. If an n-doped InSb crystal with a lower level of doping is used, a weaker magnetic bias would be required (still, several T) to obtain SPPs with reasonable propagation lengths. However, the bias and bandgap frequency range are still very much higher than for the undoped case.

 It is worth noting that properties of the metal layer also impact performance. For example, using parameters of the undoped sample in Table. II, if the metal permittivity were $\epsilon_m^\prime=(-2.3+i8.6)\times10^5$ obtained using a standard Drude model, the maximum propagation length of the SPP between two bandgaps is $7.4 \lambda_{SPP}$, more than twice the value reported in Table II. Metal deposition quality, surface roughness, etc., will also play an important role which are not the subject of this work. However, for some applications such as switches and nonlinear devices, a very long SPP propagation length is not required.

\begin{figure} 
	\includegraphics[width=1.0\columnwidth]{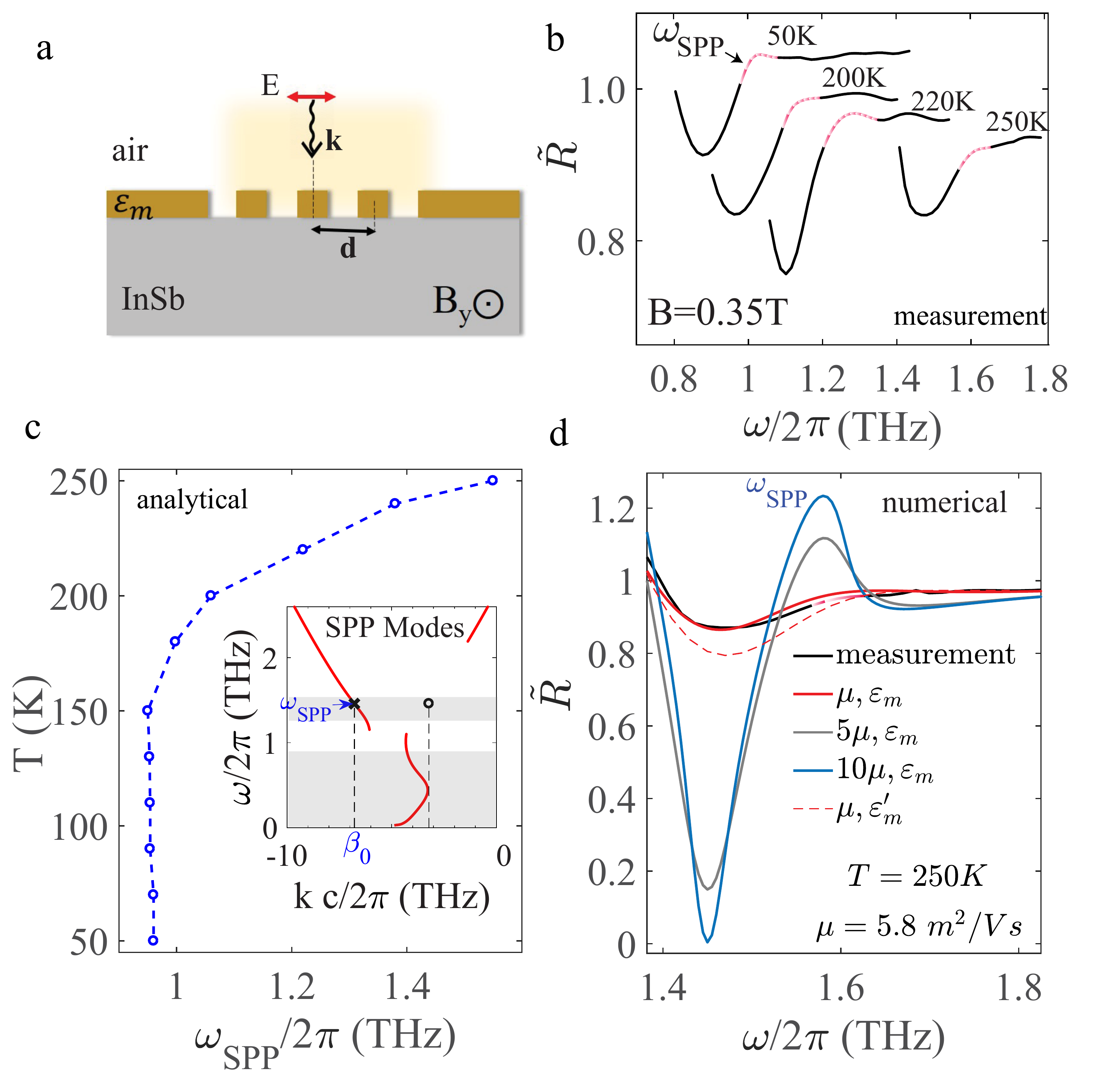}
  \caption{The far-field measurement of the unidirectional SPP at various temperatures. (a) The schematic geometry of the structure under test. The SPP is excited using a symmetric metal grating under normally incident plane wave. The period of grating is $d=84\mu \text{m}$ the metal thickness is $t=0.5\mu \text{m}$. (b) The reflection spectrum of the pattern InSb sample, measured at various temperatures. $\tilde{R}$ is defined as the ratio of r(B)/r(0) where $r(B)$ and $r(0)$ are the reflectance of the biased/unbiased pattern sample, respectively. The red peaks are observed unidirectional SPP resonances.
(c) The analytically estimated SPP resonance frequencies at different temperatures, obtained using the SPP dispersion diagram and the effective momentum $\beta_0=2\pi/d$ due to the symmetric grating as illustrated in the insert plot.
(d) The reflection spectrum of the pattern InSb, results from a COMSOL simulation. The InSb crystal is characterized by $n_e=4.5\times
10^{21}\text{m}^{-3}$, $m^{\ast}=0.0175m_{0}$, $\varepsilon_{\infty}=15.68$ and different mobility values. $\epsilon_m=-800+i550$, $\epsilon_m^\prime=(-2.3+i8.6)\times10^5$ and $B=0.35T$. 
	}
		 \label{exp}
\end{figure}

In order to excite SPPs in the experiment, a symmetric metal grating is used as the SPP launcher. Figure \ref{exp}a shows the geometry of the structure, an InSb sample covered with a grating under a normally incident plane wave. The electric field polarization is perpendicular to the grating strips as well as the magnetic bias. Figure \ref{exp}b shows the measured reflectance spectra $\tilde{R}$ of the magnetized pattern InSb sample at various temperatures. The red peaks indicate the SPP resonances of the biased sample. By applying B=0.35T, the SPP resonance frequencies are within or above the upper bandgap, depending on frequency. The period of the grating is $d=84 \mu$m with the filling factor $1/2$, providing the effective momentum $\beta_{0}=2\pi/d$ for the system under a normally incident plane wave. Using the SPP dispersion diagram the SPP resonance frequency can be obtained at different temperatures as shown in Fig. \ref{exp}c. The measured resonance frequencies are consistent with the values theoretically estimated. Figure \ref{exp}d shows the reflection spectrum of the pattern InSb obtained using a COMSOL simulation. As an example, for $T=250^\circ$K and considering a realistic value of mobility ($\mu=5.8 \text{m}^2 / \text{Vs}$) as shown the numerical result is well-matched with the measurement result. The reflectance corresponding to larger mobility has also been shown in Fig. \ref{exp}d. If the mobility were higher, the SPP resonance becomes stronger; the red resonance peaks in Fig. \ref{exp}b become larger.

Finally, in Fig. \ref{exp}d simulation results for reflectance are shown for a metal grating having permittivity $\epsilon_m^\prime=(-2.3+i8.6)\times10^5$, the standard Drude model (red-dashed curve). It can be seen that the THz-measured value \cite{ND} used in this work, $\epsilon_m=-800+i550$, provides better agreement with measurement than the higher-permittivity model, leading to confidence in this lower value. We note that the two different metal permittivity values do not significantly change the resonance frequencies shown in Fig. \ref{exp}c.

\section{Conclusion}
In this work, we studied the low THz characteristics of the SPPs on a metal-InSb interface with a realistic InSb model. SPPs in the bulk bandgaps are topological, and to design wideband devices based on topological SPPs, the propagation length and bandgap size are two important factors. For very low temperatures InSb is not a suitable platform for topological SPPs due to an extremely narrow bandgap, whereas for higher temperatures (say, above $250^\circ$K) the propagation length is not long enough due to low mobility. At temperatures between $150^\circ$K and $220^\circ$K, moderate bandgap width and ${L_\textrm{SPP}} /{\lambda_\textrm{SPP}} \gtrapprox 1$ is obtainable using undoped InSb.

\section*{Acknowledgement}

Funding for this research was provided by the National Science Foundation
under grant number EFMA-1741673.

\section*{Data Availability}

The datasets generated during and/or analyzed during the current study are available from the corresponding author on reasonable request.


\begin{thebibliography}{0}%
\makeatletter
\providecommand \@ifxundefined [1]{%
 \@ifx{#1\undefined}
}%
\providecommand \@ifnum [1]{%
 \ifnum #1\expandafter \@firstoftwo
 \else \expandafter \@secondoftwo
 \fi
}%
\providecommand \@ifx [1]{%
 \ifx #1\expandafter \@firstoftwo
 \else \expandafter \@secondoftwo
 \fi
}%
\providecommand \natexlab [1]{#1}%
\providecommand \enquote  [1]{``#1''}%
\providecommand \bibnamefont  [1]{#1}%
\providecommand \bibfnamefont [1]{#1}%
\providecommand \citenamefont [1]{#1}%
\providecommand \href@noop [0]{\@secondoftwo}%
\providecommand \href [0]{\begingroup \@sanitize@url \@href}%
\providecommand \@href[1]{\@@startlink{#1}\@@href}%
\providecommand \@@href[1]{\endgroup#1\@@endlink}%
\providecommand \@sanitize@url [0]{\catcode `\\12\catcode `\$12\catcode
  `\&12\catcode `\#12\catcode `\^12\catcode `\_12\catcode `\%12\relax}%
\providecommand \@@startlink[1]{}%
\providecommand \@@endlink[0]{}%
\providecommand \url  [0]{\begingroup\@sanitize@url \@url }%
\providecommand \@url [1]{\endgroup\@href {#1}{\urlprefix }}%
\providecommand \urlprefix  [0]{URL }%
\providecommand \Eprint [0]{\href }%
\providecommand \doibase [0]{http://dx.doi.org/}%
\providecommand \selectlanguage [0]{\@gobble}%
\providecommand \bibinfo  [0]{\@secondoftwo}%
\providecommand \bibfield  [0]{\@secondoftwo}%
\providecommand \translation [1]{[#1]}%
\providecommand \BibitemOpen [0]{}%
\providecommand \bibitemStop [0]{}%
\providecommand \bibitemNoStop [0]{.\EOS\space}%
\providecommand \EOS [0]{\spacefactor3000\relax}%
\providecommand \BibitemShut  [1]{\csname bibitem#1\endcsname}%
\let\auto@bib@innerbib\@empty
\end{thebibliography}%


\begin{thebibliography}{1}

\bibitem{davoyan} A.R. Davoyan and N. Engheta, Theory of wave
propagation in magnetized near-zero-epsilon metamaterials: evidence for
one-way photonic states and magnetically switched transparency and opacity, 
Phys. Rev. Lett. 111, 257401, 2013.

\bibitem {Mario2} M\'{a}rio G. Silveirinha, {Chern invariants for
continuous media}, Phys. Rev. B 92, 125153, 2015.

\bibitem {Mario3}M\'{a}rio G. Silveirinha,  Bulk-edge
correspondence for topological photonic continua, Phys. Rev.
B 94, 205105, 2016.

\bibitem {MH1}S. A. Hassani Gangaraj, M. G. Silveirinha, and G. W. Hanson,
Berry phase, Berry Potential, and Chern Number for Continuum Bianisotropic
Material from a Classical Electromagnetics Perspective, IEEE journal on
multiscale and multiphysics computational techniques 2, 3-17, DOI
10.1109/JMMCT.2017.2654962, 2017.

\bibitem{Soljacic2014} L.Lu, J.D. Joannopoulos, and M. Solja\v{c}i\'{c},
Topological Photonics, Nat. Photonics 8, 821-829, 2014.

\bibitem{rechtsman2013photonic} M.C. Rechtsman, J.M. Zeuner, Y. Plotnik,
Y. Lumer, D. Podolsky, F. Dreisow, S. Nolte, M. Segev, and A. Szameit,
Photonic Floquet topological insulators, {Nature} 496, 196,
2013.

\bibitem {Haldane2} F. D. M. Haldane and S. Raghu, {Possible realization
of directional optical waveguides in photonic crystals with broken
time-reversal symmetry}, Phys. Rev. Lett. 100, 013904, 2008.

\bibitem {airSPP} S. Pakniyat and A. M. Holmes and G. W. Hanson and S. A. H. Gangaraj and M. Antezza and M. G. Silveirinha and S. Jam and F. Monticone, Non-Reciprocal, Robust Surface Plasmon Polaritons on Gyrotropic Interfaces, IEEE Transactions on Antennas and Propagation 68, 3718-3729, 2020.

\bibitem{PRA} M.G. Silveirinha, S.A.H Gangaraj, G.W. Hanson, and M.
Antezza, Fluctuation-induced forces on an atom near a photonic topological
material, Phys. Rev. A 97, 022509, 2018.

\bibitem {Villa15} E. Moncada-Villa,  V. Fern\'andez-Hurtado, F. J. Garc\'{\i}a-Vidal, A. Garc\'{\i}a-Mart\'{\i}n, and J. C. Cuevas,  Magnetic field control of near-field radiative heat transfer and the realization of highly tunable hyperbolic thermal emitters, Phys. Rev. B, 92, 125418, 2015.

\bibitem{Fan16} F. Fan, Sh. T. Xu, X. H. Wang, and Sh. J. Chang Terahertz polarization converter and one-way transmission based on double-layer magneto-plasmonics of magnetized InSb, Opt. Express, 24, 126431-26443, 2016

\bibitem{Boyd} K. L.Tsakmakidis, L. Shen,  S. A. Schulz, X. Zheng, J. Upham, X. Deng, H. Altug,  A. F. Vakakis, and R. W. Boyd,  Breaking Lorentz reciprocity to overcome the time-bandwidth limit in physics and engineering,  Science, 356, 1260-1264, 2017

\bibitem {Ali18leaky} S. A. H. Gangaraj, and F. Monticone, Topologically-protected one-way leaky waves in nonreciprocal plasmonic structures, Journal of Physics: Condensed Matter 30, 104002, 2018.

\bibitem {Villa2} E. Moncada-Villa, A. I. Fern\'{a}ndez-Dom\'{i}nguez, J. C. Cuevas, Magnetic-field controlled anomalous refraction in doped semiconductors, J. Opt. Soc. Am. B 36, 935-941, 2019.

\bibitem {Hu19} H. Hu, L. Liu, X. Hu, D. Liu, D. Gao, Routing emission with a multi-channel nonreciprocal waveguide, Photon. Res., 7, 642-646, 2019.

\bibitem {Zhang19} W. Zhang, and X. Zhang, Backscattering-Immune Computing of Spatial Differentiation by Nonreciprocal Plasmonics, Phys. Rev. Applied, 11, 054033, 2019.

\bibitem{aliPWG} S. A. H. Gangaraj, B. Jin, Ch. Argyropoulos, F. Monticone, Broadband field enhancement and giant nonlinear effects in terminated unidirectional plasmonic waveguides, arXiv:2006.13393, 2020.

\bibitem{Palik} E. D. Palik, R. Kaplan, R.W. Gammon, H. Kaplan, R. F. Wallis, and J. J. Quinn, Coupled surface magnetoplasmon-optic-phonon polariton modes on InSb, Phys. Rev. B 13, 2497-2506, 1976. 

\bibitem{P2}E. D. Palik and J. K. Furdyna, Infrared and microwave magnetoplasma effects in semiconductors,  Rep. Prog. Phys. 33, 1193, 1970.

\bibitem {Han1}G.W. Hanson, S. Ali Hassani Gangaraj, A. Nemilentsau, Notes on
photonic topological insulators and scattering-protected edge states - a brief
introduction, arXiv:1602.02425, 2016.

\bibitem {Shi}S. Buddhiraju, Y. Shi, A. Song, C. Wojcik, M. Minkov, I.
Williamson, A. Dutt, S. Fan, Absence of unidirectionally propagating surface
plasmon-polaritons in nonreciprocal plasmonics, Nature Commun. 11, 674, doi:
10.1038/s41467-020-14504-9, 2020.

\bibitem {Ali}S. Ali Hassani Gangaraj, Francesco Monticone, Do truly
unidirectional surface plasmon-polaritons exist? Optica 6, 1158-1165, 2019.

\bibitem{kush} MS. Kushwaha, Plasmons and magnetoplasmons in semiconductor
heterostructures, Surface Science Reports 1, 1-416, 2001.

\bibitem{MTI}MTI Corp., see https://www.mtixtl.com for InSb crystals specification.

\bibitem {ND}S. Pandey, B. Gupta, A. Chanana, and A. Nahata, Non-Drude like behaviour of metals in the terahertz spectral range, Advances in Physics: X, Vol. 1, 2016. 

\bibitem {InSb} Y. Liang, S. Pakniyat, Y. Xiang, F. Shi, G. W. Hanson, and C. Cen,
“Tunable THz reflections induced by gapped magneto-plasmons in InSb,”
Opt. Mater. (submitted) (2020).

\end{thebibliography}
\end{document}